\documentclass[12pt,aps,prb,preprint,superscriptaddress,showpacs,amsfonts,amsmath,showkeys,floatfix]{revtex4-1}

\usepackage{graphicx,epsfig,amssymb}
\usepackage[utf8]{inputenc}
\usepackage{bm}
\usepackage{pstricks}
\usepackage{setspace}
\usepackage{graphicx}

\begin{document}
\title{Impedance Spectroscopy of SmB$_6$ single crystals}
\author{Jolanta Stankiewicz}
\affiliation{Departamento de F\'{i}sica de la Materia Condensada, Universidad  de Zaragoza, 50009-Zaragoza, Spain}
\author{Javier Blasco}
\affiliation{Instituto de Ciencia de Materiales de Arag\'on and Departamento de F\'{i}sica de la Materia Condensada, CSIC--Universidad  de Zaragoza, 50009-Zaragoza, Spain}
\author{Pedro Schlottmann}
\affiliation{Department of Physics, Florida State University, Tallahassee, FL 32306, USA}
\author{Monica Ciomaga Hatnean}
\affiliation{Department of Physics, University of Warwick, Gibbet Hill Road, Coventry CV4 7AL, UK}
\author{Geetha Balakrishnan}
\affiliation{Department of Physics, University of Warwick, Gibbet Hill Road, Coventry CV4 7AL, UK}

\date{\today}

\vspace{1cm}
\begin{abstract}
We report results from an in--plane and out--of--plane impedance study on SmB$_6$ single crystals, performed at low temperatures and over a wide frequency range. A universal equivalent circuit describes the dielectric behavior of this system across the transition, from surface to bulk dominated electrical conduction between 2 and 10 K. We identify the resistive, capacitive, and inductive contributions to the impedance. The equivalent inductance, obtained from fits to experimental data, drops drastically as the bulk starts to control electrical conduction upon increasing temperature. SmB$_6$ single crystals also show current--controlled negative differential resistance at low temperatures, which is brought about by Joule heating. This feature, in addition to inductive and capacitive contributions to the impedance, can give rise to the self--sustained voltage oscillations observed below 5 K (Stern {\it et al.}, Phys. Rev. Lett. {\bf 116}, 166603 (2016)).   
 \end{abstract}

\pacs{72.15.Qm, 75.20.Hr}
\maketitle

\section{Introduction}  

SmB$_6$ is a strongly correlated Kondo insulator that was recently found to have a robust surface state. It shows high--temperature metallic behavior which changes to an insulating one below 40 K as an energy gap $\Delta \approx$ 15--20 meV opens. This energy gap, brought in by hybridization between the localized Sm 4{\it f} states and weakly correlated (mostly) the Sm 5{\it d} states, should lead to a diverging resistance at lower temperatures. Instead, the resistance flattens below 5 K, which imply a metallic surface state. All the transport features, including the activated behavior below approximately 40 K and the resistance plateau for $T \lesssim$ 5 K, agree with the theoretical prediction that SmB$_6$ is a correlated topological insulator (TI).\cite{dzero} In this description, the bulk of SmB$_6$ is insulating, and non--trivial topological protected surface states are responsible for the low--temperature resistance saturation.\cite{tak,dzero1,alex,xu,kim1,yee,zhang,wol,phel} 

Most of the recent experimental research on SmB$_6$ shows conclusively that metallic surface states lie behind the leveling--off of its low--temperature resistivity. However, their origin and topological nature are still under debate.\cite {hlav} Furthermore, some results from specific heat, optical conductivity, and  quantum oscillations measurements indicate a bulk origin of SmB$_6$ metal--like behavior at low temperatures.\cite{wak2,lau,tan} There is currently no clear consensus on this point. The correlated state of SmB$_6$ seems only slightly affected by disorder or light doping with non--magnetic impurities.\cite{stan,jiao,eo} The robustness of the bulk gap and the protection against disorder seem to be intrinsic to this material.\cite{allen} All these features make SmB$_6$ an attractive candidate for applications. 

Most of the research on this system has been focused on its equilibrium properties. However, samarium hexaboride shows interesting transient behavior which is not well studied.  A large nonlinear resistance and an oscillating response upon biasing single crystals with a small {\it dc} current at low temperatures are two examples of such transient phenomena.\cite{kim3} A model which involves both surface and bulk states, in addition to a strong coupling between the thermal and the electrical properties, has recently been put forward to explain this response.\cite{stern,casas} Nevertheless, there are only few reports on the frequency--dependent transport properties of SmB$_6$.
 
Here, we report results from impedance spectroscopy (IS) experiments on SmB$_6$, performed at low temperatures and over a wide frequency range. By applying a single frequency voltage to the sample, the phase shift and amplitude of the resultant current can be determined. Both the magnitude (resistive) and phase shift (reactive) component of a frequency--dependent impedance yield relevant information about intrinsic and extrinsic dielectric relaxation processes. In a linear approach, the total response of the system can be described by inductive, capacitive, and resistive elements of an equivalent electrical network. These components are uniquely defined and give a clear and simple description of the underlying physical processes involved. 

The remainder of the paper is as follows. The experimental procedure is described in Section II. Results of measurements are reported and discussed in Section III, and conclusions are drawn in Section IV.

\section{Experiment}

\begin{figure}
\includegraphics*[width=12cm]{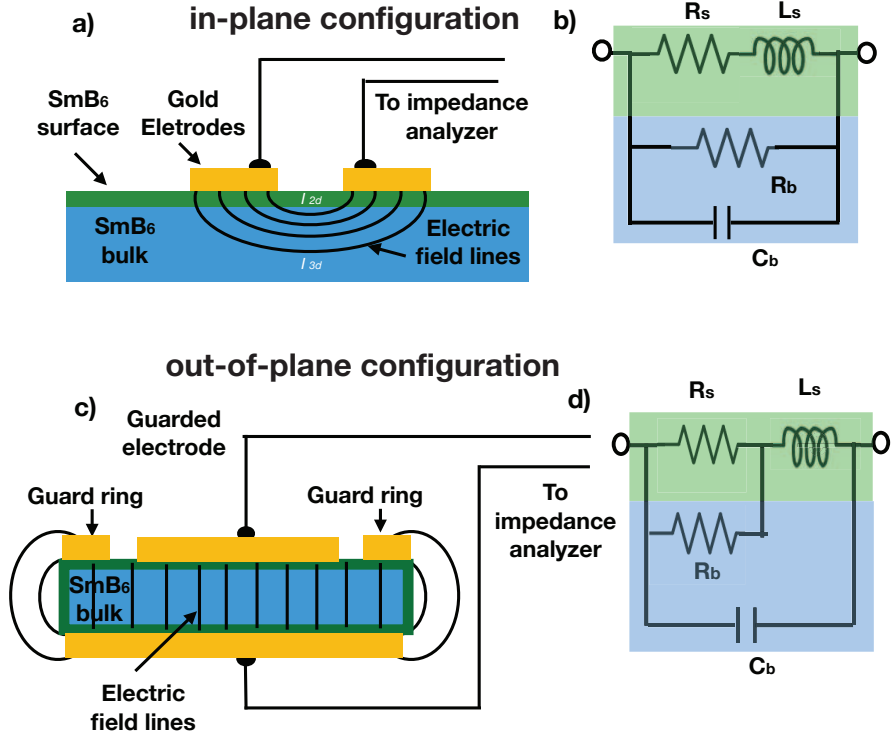}
\caption{(Color online))(a) Measurement setup for in--plane impedance spectroscopy of a SmB$_6$ single crystals in the frequency range from 20 Hz to 10 MHz and between 2 and 300 K. (b) Equivalent circuit of the device. (c) Set-up for out--of-plane impedance measurements. (d) Equivalent circuit of the device. $R_bC_b$ parallel connection models the insulating bulk contribution to the total impedance, whereas the $R_sL_s$ branch gives the inductive response from the sample surface.}
\label{Conf}
\end{figure}

Ac impedance measurements on stoichiometric and non--stoichiometric SmB$_6$ single crystals were carried out in the 1.8 to 300 K temperature range. SmB$_6$ crystals were grown by the floating zone technique using a high power xenon--arc--lamp--image furnace. This method yields large single crystals free of contamination although vacancies and defects might be still present.\cite{cio} Samples used in the experiments were selected from a batch of crystals based on their quality and size. The surfaces of these crystals, (100 planes), were carefully etched, using a mixture of hydrochloric acid and water, to remove possible oxide layers. We used silver paint, gold or pure indium to prepare contacts for impedance measurements, on recently cleaved, about 200 $\mu$m thick samples. We did not observe any important features in our measurements which could be related to contacts we have applied. Impedance spectra were obtained with a precision impedance analyzer (Wayne-Kerr) in the 20 Hz up to 10 MHz frequency range. An excitation amplitude of 0.01V  was applied to ensure the ohmic regime with a low signal--to--noise ratio. Before connecting the sample, open--circuit and short--circuit tests, in addition to high frequency compensation were carried out, to minimize parasitic contributions from the measurement setup. IS measurements were performed in both out--of plane and in--plane configurations shown in Fig.~\ref{Conf}. In the former configuration, a parallel plate capacitor configuration, stray capacitance may form on the edges of the electrodes. This causes a measurement error that can be avoided using the guard electrode which absorbs the electric field at the edge. We applied this guarded method in out--of plane studies (see Fig.~\ref{Conf}c). On the other hand, for an in--plane configuration, reducing the electrode spacing diminishes the proportion of the current carried through the bulk and increases the measurement sensitivity to the surface properties (see Fig.~\ref{Conf}a). 

To complete our study, current--voltage ($I-V$) characteristics were measured at low temperatures. The SmB$_6$ samples can be driven into a dissipative regime with increasing current below about 5 K, where the $I-V$ relation shows nonlinear negative differential resistance (NDR) from self heating. To properly measure temperature changes arising from current flow, we used a differential chromel vs. Au-0.07 at.\% Fe thermocouple between the cryostat base and the top of the sample.

\section{Results and Discussion}

Below we show and discuss the impedance spectra for several single crystalline SmB$_6$ samples. To analyze experimental data we have to ensure that the observed variations of complex impedance do arise from an intrinsic mechanism and not from the extrinsic Maxwell--Wagner type polarization effects brought about by interfaces.\cite{loid1} Most interfaces, like a Schottky barrier at the electrode--sample contact or a thin insulating surface layer with a slightly different stoichiometry, may behave as a parallel-plate capacitor with a very large capacitance and, consequently, can give rise to a very high apparent value of the dielectric constant.\cite{loid2} To rule out these effects, we measured SmB$_6$ single crystals using different contacts in various configurations. We contacted our samples with a silver conducting paste, sputtered gold, and by a direct deposition of pure In or In/Ga alloy. The variation of contact type did not lead to any significant change in impedance spectra. Therefore, we are quite confident that the reported spectra are intrinsic to SmB$_6$. We give measured quantities in absolute values. Specific values, which are necessarily determined from the bulk sample geometry, are wrong in the surface--conduction dominated region as the thickness of topologically protected metallic surface states is not reliably known. Here, we discuss data obtained for $T\lesssim$ 15 K where all interesting features happen. In the temperature range beyond 15 K, the total impedance signal, which is very small and quite noisy, corresponds to the bulk semiconductor only.

\begin{figure}
\includegraphics*[width=8cm]{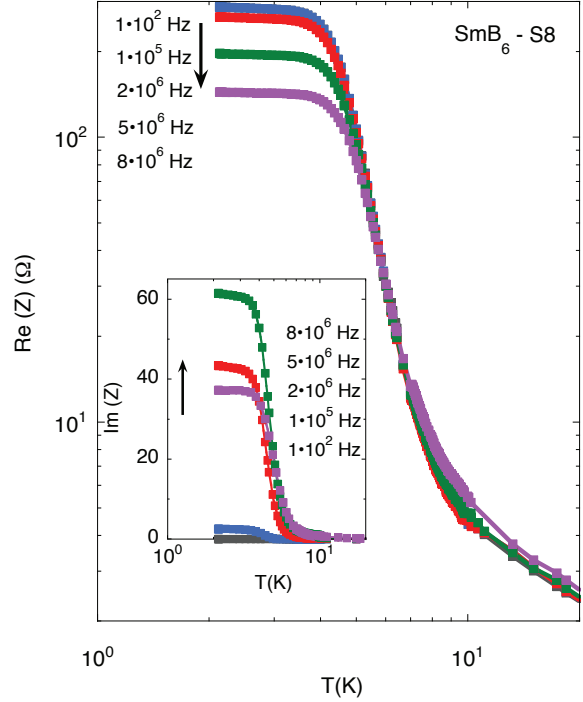}
\caption{(Color online) Real part of the in-plane impedance for single crystal SmB$_6$ (sample S8) {\it vs.} temperature. The inset shows the imaginary part of the impedance {\it vs.} temperature for the same crystal.}
\label{S8-Z-T}
\end{figure}

In Fig.~\ref{S8-Z-T}, we plot the real $Re(Z)$ and imaginary $Im(Z)$ part of impedance $Z$ for a single crystalline SmB$_6$ (sample S8) as a function of temperature $T$, in the range from 2 to 20 K, for various frequencies. These data have been obtained for the in-plane configuration, with a probe separation of approximately 150 $\mu$m. (see Fig.~\ref{Conf}a)). Upon lowering the temperature, $Re(Z)$ starts to increase quite sharply at about 20 K and levels out below $T \lesssim$ 5 K. Previous studies of high quality SmB$_6$ single crystals have shown the evolution of this system from a semi-metallic state at room temperature to a Kondo correlated insulating state, and finally to a very low carrier density metallic state at low temperatures. It is now generally accepted that, for $T \lesssim$ 5 K, the transport in SmB$_6$ takes place through surface states (SS), as the bulk resistance $R_b$ is much higher than the surface resistance $R_s$. With increasing temperature (4 $\lesssim T \lesssim$ 10 K), $R_b$ becomes comparable to that of the SSs, and conduction proceeds through both the bulk of the crystal and the surface channels. Beyond 10 K and up to approximately 40 K, the bulk resistance over this temperature range is much smaller than that of the surface states, and the latter have no effect on electrical transport. The data plotted in Fig.~\ref{S8-Z-T} follow this scheme. Assuming that the surface conductivity is independent of temperature, we obtain for the bulk resistance a thermally activated behavior $R_b \propto exp(\Delta E/k_B T)$, with an activation energy of 3.3 meV, for 2 K$< T < $10 K. 

The imaginary part of $Z(T)$ for the sample S8 also shows a pronounced rise below 5 K for large excitation frequencies, followed by leveling out at lower temperatures (see inset in Fig.~\ref{S8-Z-T}). The Kondo insulator SmB$_6$ appears to have an intrinsic metal to high-dielectric material transition. As its insulating gap opens completely at low temperatures, the bulk becomes an insulator with a large dielectric constant, whereas the surface states are metallic. This can be represented by an equivalent parallel $RC$ circuit and a topological surface state.\cite{dzero} 

\begin{figure}
\includegraphics*[width=15cm]{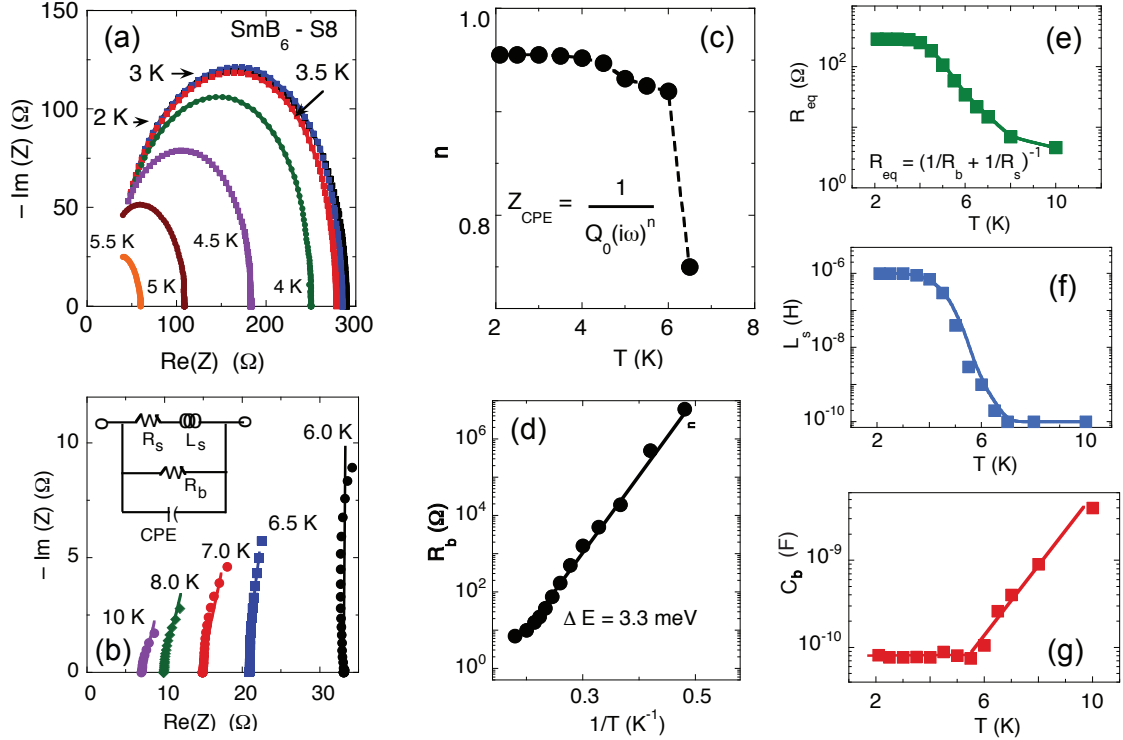}
\caption{(Color online) (a) and (b) - IS data for $T \leq $ 10 K,  plotted as the negative {\it vs.} real part of the impedance (Cole-Cole plot) for SmB$_6$ crystal (sample S8). Symbols indicate the data; solid lines represent the fit of the data to the model shown in Fig.~\ref{Conf} (b). (c) - variation of the constant phase element exponent $n$ with temperature. (d) through (g) - fitted temperature dependences of the equivalent circuit elements for SmB$_6$ (sample S8): (d) - $R_b$ as a function of $1/T$. (e) $R_{eq}$, as defined in the figure. (f) Inductance $L_s$, showing a large drop between 4 and 6 K. (g) Capacitance $C_b$.}
\label{S8}
\end{figure} 

Figures~\ref{S8}(a) and (b) show the IS data plotted as the negative imaginary $-Im(Z)$ vs. the real part $Re(Z)$ (Cole--Cole plot) at various temperatures in the range from 2 to 10 K. In such impedance plots, one or more semicircles may be seen for a parallel resistor-capacitor $RC$ circuit measurement; each semicircle corresponding to a dielectric relaxation arising from different, extrinsic (interfaces) or intrinsic (sample), contributions in series.\cite{bor} Our data show one single, non-ideal dielectric relaxation, evidenced by one depressed semicircle for each temperature, where the center of each semicircle has shifted below the x-axis ($Re(Z)$). For a Debye--like relaxation (an ideal parallel resistor-capacitor element), a perfect semicircle is expected.\cite{jon} Since the data cannot be modeled solely using standard $RC$ elements, we used the alternative equivalent circuit depicted in Fig.~\ref{Conf}(b), which yields good fits at all frequencies. The ``depressed" behavior can be fitted by adding a series $R_sL_s$ branch, in parallel to the conventional $RC$ element. Physically, the $R_bC_b$ part accounts for the insulating SmB$_6$ bulk, with a standard dielectric behavior. (Fig.~\ref{Conf}(b)). The $R_sL_s$ branch corresponds to the inductive response, most likely from the sample surface. We also examined the equivalent circuit, displayed in the inset of Fig.~\ref{S8}(b),\cite{jon} which gives a somewhat better fit for the sample S-8 data. The Constant Phase Element ($CPE$) has an impedance of the form $Z(CPE) = 1/(Q_o(i\omega)^n)$ where $Q_o$ models the effects of a non-ideal relaxation process or retardation. The exponent $n$ is a measure of the non-ideality of the relaxation process, brought about by non-uniformity of physical properties of the system.  When $n$ = 1, the system is described by a single time-constant and the parameter $Q_o$ has units of capacitance. Its variation with $T$, obtained from fittings to the circuit in the inset of Fig.~\ref{S8}(b), is shown in Fig.~\ref{S8}(c).

The fitting parameters we have extracted from our model for the S8 sample are shown in Figs.~\ref{S8} (d)--(g), where each point on the curves represents a fit at a specific temperature. For this sample, $R_s = 288.5 \Omega $. As discussed above, the bulk resistance $R_b$ follows thermally activated behavior with an activation energy of 3.3 meV for 2 K$ < T < $10 K. The equivalent sample resistance $R_{eq}$ is calculated as $(1/R_b +1/Rs)^{-1}$. It is worth noticing that $R_b$ and $R_{eq}$ are very close above 6 K, suggesting that the current flows mainly through bulk beyond this temperature.  Fig.~\ref{S8}(g) depicts the sample capacitance $C_b$ we have estimated from the model. The capacitance stays constant as the temperature increases up to about 6 K, where it upturns. For the in-plane configuration (see Fig.~\ref{Conf}(a)), the field lines of the applied {\it ac} electric field are not parallel in the sample and the changes in capacitance may not necessarily reflect intrinsic changes in the bulk dielectric permittivity.  
Figure~\ref{S8}(f) shows temperature variation of fitted values for the inductance $L_s$. In the low--temperature metallic state below 4K, $L_s$ $\approx$ 10$^{-6}$ H. As the temperature increases, $L_s$ drops by four orders of magnitude between 4 and 6 K, and is negligibly small farther on.  We note that the temperature variations of $L_s$ and $R_{eq}$ are alike in the temperature range considered. The large drop of $L_s$ occurs at the transition from surface to bulk dominated conduction. The inductive contribution in the metallic regime may arise from the coupling of 2D (metallic) and 3D (bulk) conduction channels, likely through defects.\cite{jash}. On the other hand, recent scanning microwave impedance microscopy results show presence of one--dimensional conducting channels terminating at surface step edges.\cite{xuBarber} These conducting channels may contribute significantly to the total inductance at low temperatures. 

\begin{figure}
\includegraphics*[width=8cm]{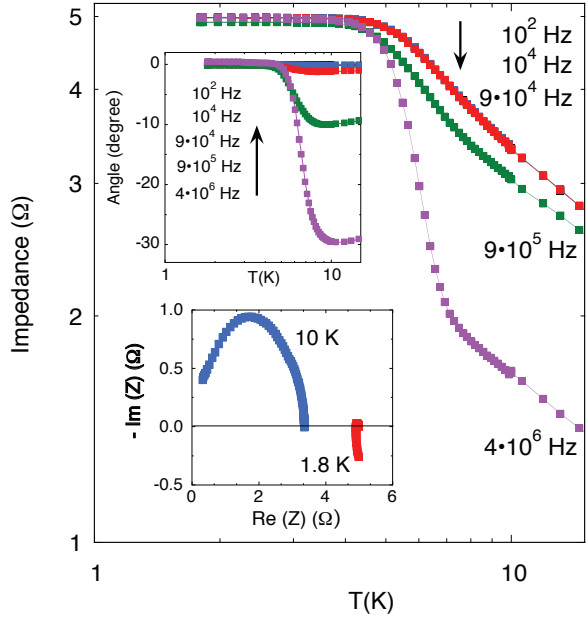}
\caption{(Color online) In-plane impedance for single crystal SmB$_6$ (sample S9) as a function of temperature at various frequencies. The upper inset shows phase angle {\it vs.} temperature. The lower inset displays the Cole--Cole plot at 1.8 and 10 K for the same sample. Note that $Im(Z)$ is positive at 1.8 K. }
\label{S9-all}
\end{figure}

The IS data for SmB$_6$ (sample S9) are reported in Fig.~\ref{S9-all}. This sample is identical to the S8 sample, except for the contact separation (approximately 25 $\mu$m) which is almost one order of magnitude smaller than in sample S8. Therefore, we expect surface effects to be more pronounced. Indeed, the total impedance increase at high frequencies shows behavior typical of inductive contributions. The imaginary impedance $Im(Z)$ is then positive at 1.8 K because the spectra are dominated by the surface inductance $L_s$. The model applies just the same.

\begin{figure}
\includegraphics*[width=15cm]{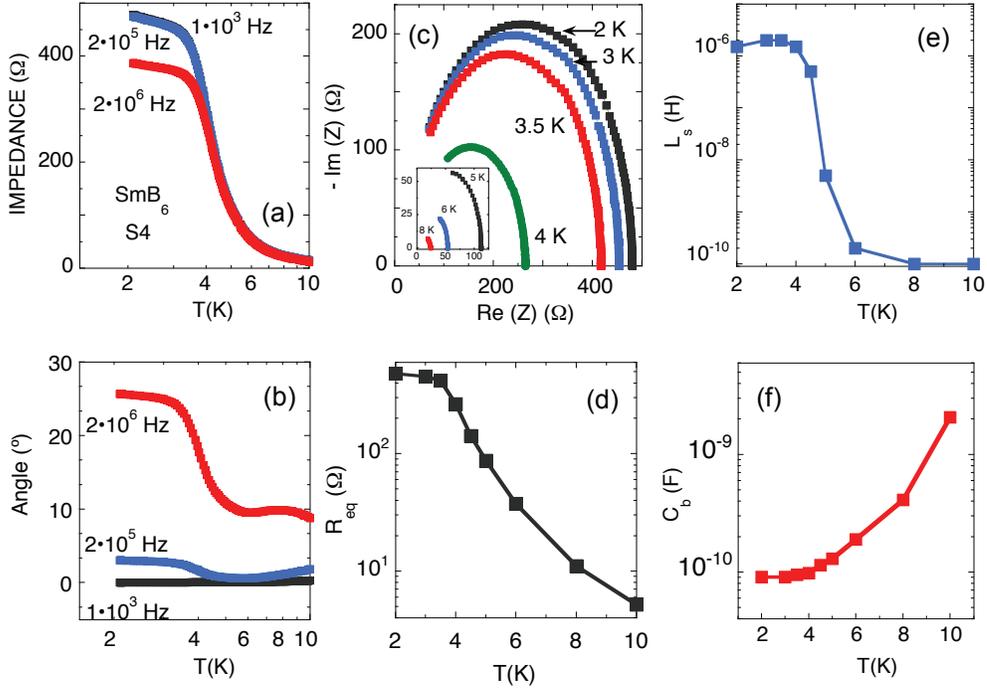}
\caption{(Color online) (a)- Guarded out-of plane impedance for single crystal SmB$_6$ (sample S4) as a function of temperature at various frequencies. (b)- Phase angle $vs.$ temperature for the same sample. (c)- Cole--Cole plot between 2 and 10 K. (d)- $R_{eq}$, obtained from the fit to the equivalent circuit {\it vs.} $T$. (e)- Values of the fitted inductance $L_s$, and (f)- capacitance $C_b$ as a function of $T$.}
\label{S4}
\end{figure}

The IS results obtained for the out-of plane configuration (see Fig.~\ref{Conf}(b)) of SmB$_6$ single crystals (samples S1--S4) are much the same as the results for the in-plane configuration. Figure~\ref{S4}(c) shows the Niquist plot for sample S4 at several temperatures in the range 2 K-10 K. Variations of the magnitude of the impedance $Z(T)$ and of the phase angle $\theta(T)$ at various frequencies are displayed in Figs.~\ref{S4}(a) and (b). The values of the parameters, obtained from the fit to the equivalent circuit of Fig.~\ref{Conf}, are plotted in Figs.~\ref{S4}(d)--(f). Including the $CPE$ element in the equivalent circuit instead of $C_b$ does not improve our fittings. As discussed above, the inductive contribution is important below 4 K, drops by several orders of magnitude with increasing temperature and becomes negligible beyond 6 K. 

Our fits to experimental data were carried out with the least--square method. The relative error in parameters $L_s$ and $C_b$ is less than 5\% for $T \lesssim$ 5 K.  However, for $T \ge 6 K$, the fitting parameters may be more uncertain because of the small signal measured. In addition, we observe some parasitic stray inductance for the sample S8 (in--plane configuration) which bends curves in the Cole--Cole plot very slightly to the right for $T \ge 6 K$. Since any contribution to the impedance from the experimental setup (wires plus apparatus) should be temperature independent, this is brought, most likely, by electrodes.

\begin{figure}
\includegraphics*[width=8cm]{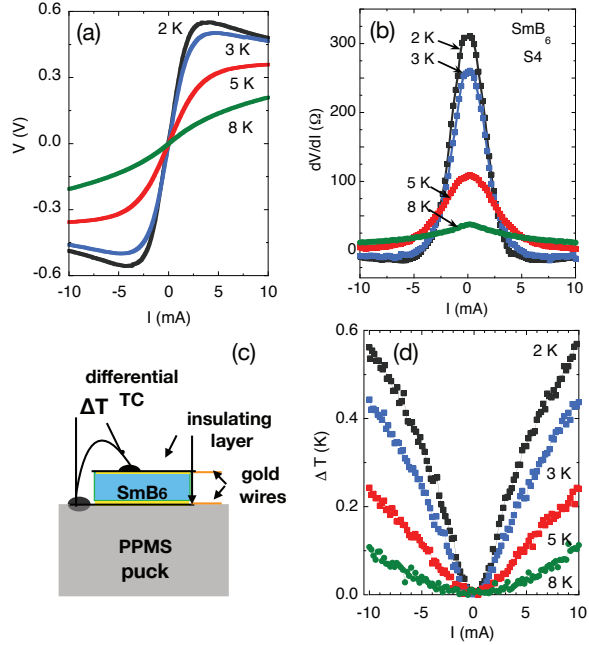}
\caption{(Color online)(a)- $I-V$ curves for single crystal SmB$_6$ (sample S4) at various temperatures. (b)- $dV/dI$ curves at the same temperatures. (c)- set--up used to measure $\Delta T$. (d)- self heating effect of $dc$ current ($\Delta T~ vs.~ I$) measured for the same sample.}
\label{I-V}
\end{figure}

In Fig.~\ref{I-V}, we plot current-voltage ($I-V$ and $dV/dI$) characteristics measured at low temperatures for sample S4. $I-V$ relation shows nonlinear negative differential resistance (NDR) with increasing current below about 5 K. Figure~\ref{I-V}(d) displays the change in sample temperature corresponding to this process. The observed increment in temperature, $\Delta T$, is less than 0.6 K over the base temperature of 2 K and is brought about by self heating. Similar behavior, albeit with a larger $\Delta T$, for SmB$_6$ single crystals has been reported in Ref. \onlinecite{kim3}. In addition, the authors of this report find large voltage oscillations driven by a small {\it dc} current. To account for this, a model, in which a coupling between metallic surface conduction and thermally activated bulk conduction giving rise to a large variation in sample heating, was proposed.\cite{stern} Qualitatively, the Joule heating in the surface conduction regime rises locally the sample temperature. As a result, the sample resistance decreases, enabling bulk conduction, and thus reducing Joule self heating. Consequently, the sample temperature drops to the initial value and drives the system back to the surface conduction regime. This model, for a rather narrow range of the parameters, yields voltage oscillations that agree with experiment.

In our set--up, we do not find self-sustained oscillations in the SmB$_6$ crystals. The dynamics of the model reported in Ref. \onlinecite{stern} depends importantly on the ratio of the bulk resistance to the surface resistance in thermal equilibrium. This ratio is about or less than 80 in the crystals of Refs. \onlinecite{kim3}, \onlinecite{stern} and \onlinecite{casas}, whereas it is larger than $10^4$ for the SmB$_6$ samples we have studied. As a result, raising the sample temperature by 1 K in the former case (from a base temperature of 2 K) is sufficient to drive SmB$_6$ crystals into a bulk conduction regime but our samples must be heated at least 5 K in order to reach the same conduction regime. On this account, the sample heating upon biasing with a {\it dc} current is much smaller in our experiments.  In addition, the model of Ref. \onlinecite{stern} involves local effects and it is not known which portion of the sample is affected by current. It is not so in the experimental configuration we have used in which the whole sample is subjected to heating. These features might have hindered the non-linear oscillations found in Ref. \onlinecite{stern} as well as the shape (surface area and thickness) of the SmB$_6$ platelets. 

On the other hand, it is well known that many dissipative systems, such as inductor-capacitor-resistor ($LCR$) circuits with a nonlinear negative--differential resistance show self-sustained oscillations which can be described by the van der Pol equation.\cite{rand,pol} As shown above, the SmB$_6$ low-temperature impedance spectra can be interpreted by an equivalent $LCR$ circuit, with NDR coming from the self heating. It is therefore reasonable to ask whether the self-sustained voltage oscillations in SmB$_6$ are the ones depicked by the van der Pol equation. If so, the oscillation frequency should be inversely proportional to the square root of the effective capacitance times inductance, i.e., $LC \propto 1/\omega^2$. This verifies that SmB$_6$ crystals may work as a current-controlled oscillator in the 20 MHz frequency range as shown in Ref.~\onlinecite{stern}.

\section{Conclusions}

In-plane and out-of-plane impedance spectra were measured in SmB$_6$ single crystals in order to study the variation of the dielectric response of this system at low temperatures. A universal model describes quantitatively the complex impedance data in surface- and bulk--conduction dominated regimes. The model includes a $R_bC_b$ circuit and an inductive $R_sL_s$ branch in parallel. The latter becomes dominant at high frequencies. The $R_bC_b$ element is made of a resistor and a capacitor connected in parallel. It
describes standard dielectric relaxations in bulk. The $R_sL_s$ branch accounts for the inductive contribution seen below 5 K, in the surface--conduction metallic regime. The equivalent inductance, obtained form fits to experimental data, drops drastically with increasing temperature as the bulk starts to control electrical conduction. We attribute this inductive contribution to the surface states and its coupling to the bulk at low temperatures. Platelet shaped SmB$_6$ single crystals show a current--controlled negative--differential resistance at low temperatures, which is brought by a Joule heating. The inductive and capacitive contributions to the impedance, obtained for SmB$_6$ single crystals at low temperatures, can give rise to the self--sustained voltage oscillations found in Ref.~\onlinecite{stern}.

We acknowledge the financial support of Ministerio de Ciencia, Innovaci\'{o}n y Universidad of Spain through Grant RTI2018-098537-B-C22 co--funded by ERDF, EU.
The work at the University of Warwick was supported by EPSRC, UK, through Grant EP/T005963/1.

\end{document}